\documentclass[12pt]{article}
\usepackage{amsmath}
\usepackage{amsthm}
\usepackage{graphicx}
\usepackage{txfonts}
\usepackage{enumerate}
\usepackage[colorinlistoftodos]{todonotes}
\usepackage[a4paper, margin=1in]{geometry}
\usepackage{amssymb}
\usepackage{apacite}
\usepackage{listings}

\usepackage{xcolor}
\usepackage[colorlinks = true,
            linkcolor = blue,
            urlcolor  = blue,
            citecolor = blue,
            anchorcolor = blue]{hyperref}
\usepackage{float}
\usepackage{bm}
\usepackage{natbib}
\usepackage{times}
\usepackage{caption}
\usepackage{subcaption}
\usepackage[plain,noend]{algorithm2e}
\usepackage{float}
\newtheorem{theorem}{Theorem}
\numberwithin{theorem}{section}

\numberwithin{Lemma}{section}

\numberwithin{Definition}{section}

\allowdisplaybreaks
\usepackage{fancyhdr}
\fancypagestyle{plain}{%
    \fancyhf{}%
    \fancyfoot[R]{\thepage}%
}
\usepackage{color,soul}
\usepackage{array}
\newcolumntype{L}[1]{>{\centering}m{#1}}
\usepackage{float}
\floatstyle{plaintop}
\restylefloat{table}
\usepackage{multirow}
\usepackage{longtable}
\begin{document}
\title{ \large Nonparametric Confidence Intervals for Generalized Lorenz Curve Using Modified Empirical Likelihood}
\author{\small Suthakaran Ratnasingam $^\dag$, Spencer Wallace$^*$, Imran Amani$^*$, Jade Romero$^*$ \\
\small  Department of Mathematics \\ \small California State University, San Bernardino, San Bernardino, CA 92407, USA}
\date{}
\maketitle
\def\thefootnote{\dag}\footnotetext{Corresponding author. Email: suthakaran.ratnasingam@csusb.edu}\def\thefootnote{\arabic{footnote}}
\def\thefootnote{*}\footnotetext{These authors contributed equally to this work}\def\thefootnote{\arabic{footnote}}
\vspace{-1cm}
\begin{abstract}
\noindent

\sloppy The Lorenz curve portrays income distribution inequality. In this article, we develop three modified empirical likelihood (EL) approaches, including adjusted empirical likelihood, transformed empirical likelihood, and transformed adjusted empirical likelihood, to construct confidence intervals for the generalized Lorenz ordinate. We demonstrate that the limiting distribution of the modified EL ratio statistics for the generalized Lorenz ordinate follows scaled Chi-Squared distributions with one degree of freedom. We compare the coverage probabilities and mean lengths of confidence intervals of the proposed methods with the traditional EL method through simulations under various scenarios. Finally, we illustrate the proposed methods using real data to construct confidence intervals.

\end{abstract}

\section{Introduction}

\sloppy The Lorenz curve developed by American economist Max Lorenz (\cite{lor1905}) is a graphical representation used to describe income and wealth inequality. A Lorenz curve with perfect equality follows a diagonal line (45$^{\circ}$ angle) in which the income percentage is always proportional to the population percentage; however, in the real world, the Lorenz curve falls below this line. As the actual income distribution is rarely known, the distribution is typically estimated from income data. Several researchers have made contributions to Lorenz curves analysis, for example, \cite{sen1973}, \cite{jak1976}, \cite{goldie1997}, and \cite{marsh1979}. The full joint variance-covariance structure for the Lorenz curve ordinates was developed by \cite{beach1983}. \cite{john1989} proposed new results on generalized Lorenz ordinate analysis that are relevant for proving second-degree stochastic dominance. \cite{all1990}, \cite{Lambert2001}, and \cite{Mosler2007} have made recent advances, with their findings leading to a wide range of applications, particularly in reliability theory. \cite{ryu1996} proposed an exponential polynomial expansion and a Bernstein polynomial expansion as two flexible functional form approaches for approximating Lorenz curves. \cite{hase2003} proposed a Bayesian non-parametric analysis approach with the Dirichlet process prior to Lorenz curve estimation with contaminated data. In particular, their method allows for heteroscedasticity in individual incomes. Further, the Lorenz curve has been used by several researchers to analyze physician distributions. For example, \cite{chang1997} examined variations in the distribution of pediatricians among the states between 1982 and 1992 using Lorenz curves and Gini indices. \cite{kob1992} used the Lorenz curve and the Gini coefficient to study the disparity in physician distribution in Japan.\\

\sloppy Empirical likelihood (EL) is a nonparametric method introduced by \cite{owen2001}, an alternative to the standard parametric likelihood that inherits many alluring features such as its extension of Wilks' theorem, asymmetric confidence interval, better coverage for small sample sizes, and so on. Many researchers have studied EL for the Lorenz curve. For instance, \cite{belinga2007} and \cite{yang2012} developed plug-in empirical likelihood-based inferences to construct confidence intervals for the generalized Lorenz curve. \cite{qin2013} studied EL-based confidence interval for the Lorenz curve under the simple random sampling and the stratified random sampling
designs. \cite{shi2019} proposed new nonparametric confidence intervals using the influence function-based empirical likelihood method for the Lorenz curve and showed that the limiting distributions of the
empirical log-likelihood ratio statistics for the Lorenz ordinates were standard chi-square
distributions. \cite{luo2019} suggested a kernel smoothing estimator for the Lorenz curve and developed a smoothed jackknife empirical likelihood approach for constructing confidence intervals of Lorenz ordinates.\\

\sloppy Despite being widely used, the EL-based approach has two major drawbacks: (1) The convex hull must have vector zero as its interior point in order to solve the profile empirical likelihood problem. According to \cite{owen2001}, the empirical likelihood function should be set to $-\infty$ if the convex hull does not have zero as an interior point. However, \cite{chen2008}, pointed out that this makes it difficult to find the maximum of the EL function. (2) The EL technique frequently experiences under-coverage problems, see  \cite{tas2013} for more details. To address these problems, numerous strategies have been proposed in the literature. \cite{chen2008} proposed adjusted empirical likelihood method (AEL), for example, confirms the existence of a solution in the maximization problem while preserving asymptotic optimality properties. Further, \cite{jing2017} suggested the transformed empirical likelihood to tackle the under-coverage problem for small sample sizes (TEL). \cite{pat2020} proposed the transformed adjusted empirical likelihood (TAEL), a strategy that combines the AEL and TEL approaches. The AEL, TEL, and TAEL approaches are proven to be effective in many applications, for example, \cite{li2022} investigated modified EL-based confidence intervals for quantile
regression models with longitudinal data.  \cite{rat2020} studied all three modified versions of EL procedures to construct confidence intervals of the mean residual life function with length-biased data.  \\

In this research, we develop three modified EL-based inference procedures to construct confidence intervals for the generalized Lorenz curve. These modified EL methods aim to address the shortcomings of traditional EL, including the issue of under-coverage, while also ensuring the existence of a solution for the maximization procedure. To the best of our knowledge, this is the first study to investigate AEL, TEL, and TAEL methods for constructing confidence intervals for the generalized Lorenz ordinate.\\

The remainder of this paper is organized as follows. In Section 2, we briefly describe the fundamental properties of EL for the generalized Lorenz curve and provide the methodology of AEL, TEL, and TAEL for the generalized Lorenz curve. In Section 3, we conduct an extensive simulation study to compare the finite sample performances of the proposed confidence intervals for the generalized Lorenz ordinates. In Section 4, we use an income dataset to illustrate the proposed intervals. In Section 5, we discuss our results and draw conclusions. 


\section{Empirical Likelihood  Based Methods}

\subsection{Empirical Likelihood}
Let $X$ be a random variable with cumulative distribution function (CDF) denoted by $F(x)$ with finite support. For instance, $F(\cdot)$ denotes the CDF of income or wealth distribution.
Following \cite{gas1971}, a general definition of the Lorenz curve is provided below.
\begin{equation}\label{}
\begin{aligned}
\eta(t) = \dfrac{1}{\mu} \int_{0}^{\psi_{t}}x dF(x), \quad t  \in [0,1]
\end{aligned}
\end{equation}
where $\mu$ denotes the mean of $F$, and $\psi_{t} = F^{-1} (t)  = \inf \{x : F(x) \geq t\}$ is the $t-$th quantile
of $F$. For a fixed $t \in [0, 1]$, the Lorenz ordinate $\eta(t)$ is the ratio of the mean income of the lowest $t$-th fraction of households and the mean income of total households. The generalized Lorenz curve is defined as follows.
\begin{equation}\label{}
\begin{aligned}
\theta(t) = \int_{0}^{\psi_{t}}x dF(x), \quad t  \in [0,1]
\end{aligned}
\end{equation}
Because the income distribution $F(x)$ is rarely known in practice, the Lorenz curve is typically estimated from income data. Hence, the empirical estimator for $\eta(t)$ is defined as
\begin{equation}\label{}
\begin{aligned}
\widehat{\eta}(t) = \dfrac{1}{\widehat{\mu}} \int_{0}^{\widehat{\psi}_{t}}x d\widehat{F}_{n}(x), \quad t  \in [0,1]
\end{aligned}
\end{equation}
where $\widehat{F}_{n}(x)$ is the empirical distribution function of the $X_{i}$'s, $\widehat{\mu}$ is the sample mean, $\widehat{\psi}_{t}$ is the $t-$th sample quantile of the $X_{i}$'s. From the definition of the generalized Lorenz curve, we observe that 
\begin{equation*}
\begin{aligned}
E[ X I(X \leq \psi_{t})] - \theta(t) = 0.
\end{aligned}
\end{equation*}
Therefore, the empirical likelihood of $\theta(t)$ can be expressed as
\begin{equation}\label{sep8}
\begin{aligned}
L^{*}(\theta(t)) = \sup_{p}\bigg\{\prod_{i=1}^{n}p_{i}: \sum_{i=1}^{n}p_{i}=1, \sum_{i=1}^{n}p_{i} W_{i}(t) = 0 \bigg\},
\end{aligned}
\end{equation}
where $\mathbf{p} = (p_{1}, p_{2}, \dots, p_{n})$ is a probability vector satisfying $\sum_{i=1}^{n}p_{i} = 1$ and $p \geq 0$ for all $i$, and $W_{i}(t) = X_{i}I(X_{i}\leq \psi_{t}) - \theta(t), i = 1,2, \dots,n$. It can be seen that $W_{i}(t)$ in (\ref{sep8}) depends on the unknown $t-$quantile $\psi_{t}$. As a result, the generalized Lorenz ordinate $\theta(t)$ is the mean of the random variable $X$ truncated at $\psi_{t}$. Using sample data, the empirical likelihood for $\theta(t)$ as follows: 
\begin{equation}\label{sep81}
\begin{aligned}
L(\theta(t)) = \sup_{p}\bigg\{\prod_{i=1}^{n}p_{i}: \sum_{i=1}^{n}p_{i}=1, \sum_{i=1}^{n}p_{i} \widehat{W}_{i}(t) = 0 \bigg\},
\end{aligned}
\end{equation}
where $\widehat{W}_{i}(t) = X_{i}I(X_{i}\leq \widehat{\psi}_{t}) - \theta(t), i = 1,2, \dots,n$. When the vector $\mathbf{p} = (p_{1}, p_{2}, \dots, p_{n})$  is contained within the convex hull defined by $\{X_{1}I(X_{1}\leq \widehat{\psi}_{t}), X_{2}I(X_{2}\leq \widehat{\psi}_{t}), \dots, X_{n}I(X_{n}\leq \widehat{\psi}_{t}) \},$ equation (\ref{sep81}) attains its unique maximum value.  By applying the Lagrange multiplier method, we can determine, $L(\theta(t))$ as follows.
\begin{equation*}
\begin{aligned}
p_{i} = \dfrac{1}{n} \bigg\{ 1 + \lambda(t) \widehat{W}_{i}(t)\bigg\}^{-1}, \quad i = 1,\dots, n.
\end{aligned}
\end{equation*}
where $\lambda$ is the solution to
\begin{equation*}
\begin{aligned}
\dfrac{1}{n} \sum_{i=1}^{n} \dfrac{\widehat{W}_{i}(t)}{1 + \lambda(t) \widehat{W}_{i}(t)} = 0.
\end{aligned}
\end{equation*}
Note that $\prod_{i=1}^{n} p_{i}$, subject to $\sum_{i=1}^{n}p_{i} = 1$, attains its maximum $n^{-n}$ at $p_{i} = n^{-1}$. Thus, the EL ratio for $\theta(t)$ is given as
\begin{equation}\label{}
\begin{aligned}
\mathcal{R} (\theta (t)) = \prod_{i=1}^{n}np_{i} = \prod_{i=1}^{n} \big\{1 + \lambda(t) \widehat{W}_{i}(t) \big\}^{-1}.
\end{aligned}
\end{equation}
Hence, the profile empirical log-likelihood ratio for $\theta(t)$ is
\begin{equation}\label{eqaug3022}
\begin{aligned}
\ell(\theta(t)) = -2 \log \mathcal{R} (\theta (t)) = 2 \sum_{i=1}^{n}\log \big\{1 + \lambda(t) \widehat{W}_{i}(t) \big\}.
\end{aligned}
\end{equation}

\begin{theorem}\label{thm0}
If $E(X^{2})<\infty$ and $\theta(t_{0}) = E[ X I(X \leq \psi_{t_{0}})]$ for any given $t = t_{0} \in (0,1)$ then the limiting distribution of $\mathrm{}{l}(\theta(t_{0}))$ is a scaled chi-square distribution with one degree of freedom,
\begin{equation}\label{}
\begin{aligned}
\bigg(\dfrac{\sigma_{p}^{2}}{\sigma^{2}_{v}}\bigg)\ell(\theta(t_{0})) \longrightarrow \chi^{2}_{1}, \quad \text{as}\,\, n \longrightarrow \infty.
\end{aligned}
\end{equation}
where $\sigma_{p}^{2} = Var\big(X\,I(X \leq \psi_{t}) \big)$ and $\sigma_{v}^{2} = Var\big((X - \psi_{t})I(X \leq \psi_{t}) \big)$.
\end{theorem}
Although the scale constant $\bigg(\dfrac{\sigma_{p}^{2}}{\sigma^{2}_{v}}\bigg)$ is unknown, it can be consistently estimated by using the following formula.
\begin{equation}\label{}
\begin{aligned}
\sigma_{p}^{2} &= \dfrac{1}{n}\sum_{i=1}^{n}\bigg(X_{i}\, I(X_{i} \leq \widehat{\psi}_{t}) -\dfrac{1}{n} \sum_{i=1}^{n} X_{i}\, I(X_{i} \leq \widehat{\psi}_{t}) \bigg)^{2}, \quad \text{and} \\
\sigma_{v}^{2} &=\dfrac{1}{n}\sum_{i=1}^{n}\bigg((X_{i}- \widehat{\psi}_{t}) I(X_{i} \leq \widehat{\psi}_{t}) -\dfrac{1}{n} \sum_{i=1}^{n} (X_{i}-\widehat{\psi}_{t}) I(X_{i} \leq \widehat{\psi}_{t}) \bigg)^{2}.
\end{aligned}
\end{equation}

Thus, an asymptotic $(1-\alpha)100\%$ confidence interval for generalized Lorenz ordinate, $\theta(t)$ at a fixed time~$t$~is given as follows
\begin{equation}\label{}
\begin{aligned}
C(t)=\bigg\{\theta(t):\bigg(\dfrac{\sigma_{p}^{2}}{\sigma^{2}_{v}}\bigg)\ell(\theta(t))\le \chi^{2}_{1,\alpha}\bigg\},
\end{aligned}
\end{equation}
where $\chi^{2}_{1,\alpha}$ is the upper $\alpha-$quantile of the distribution of $\chi^{2}_1$. For more details, we refer to \cite{yang2012}. As previously stated, the original EL method experiences low coverage probability, particularly for small sample sizes, for example, \cite{chen2008} and \cite{jing2017}. Next, we describe the technical details of three modified EL-based methods for constructing confidence intervals. These methods are called AEL, TEL, and TAEL, and they are used for constructing confidence intervals for the generalized Lorenz ordinate $\theta(t)$, at a fixed time $t$ for $t \in (0, 1)$.

\subsection{Adjusted Empirical Likelihood for Generalized Lorenz Ordinate}

\sloppy \cite{chen2008} proposed the adjusted empirical likelihood (AEL) in order to address the challenge of the non-existence of a solution in the equation (\ref{eqaug3022}). We adopted the idea of the AEL method for generalized Lorenz ordinate. We  define~$\overline{W}_{n}=(1/n)\sum_{i=1}^{n}W_i(t)$~. The pseudo value~$W_{n+1}(t)=-a_n\overline{W}_{n},$~where $a_n=\max \big\{1,\frac{1}{2}\log n\big\}$. Using the $(n + 1)$ observations, we define the adjusted empirical likelihood as
\begin{equation}\label{}
\begin{aligned}
L^{*}(\theta(t)) &= \sup_{p}\bigg\{\prod_{i=1}^{n+1}p_{i}\bigg| p_{i} \geq 0, \sum_{i=1}^{n+1}p_{i}=1, \sum_{i=1}^{n+1}p_{i} \widehat{W}_{i}(t) = 0 \bigg\} 
\end{aligned}
\end{equation}
Thus, the adjusted empirical log-likelihood ratio is given by
\begin{equation}\label{eq12}
\begin{aligned}
\ell^{*}(\theta(t))=2\sum_{i=1}^{n+1}\log \big(1+\lambda(t)\widehat{W}_i(t)\big).
\end{aligned}
\end{equation}
\begin{theorem}\label{thm1}
Assume that~$E(X^{2})<\infty$. For all~$t_{0}\in (0,1)$, let $\ell^{*}(\theta(t_{0}))$ be the adjusted log-empirical likelihood ratio function defined by (\ref{eq12})  and $a_{n} = o_{p}(n^{2/3})$. We have
\begin{equation}\label{}
\begin{aligned}
\bigg(\dfrac{\sigma_{p}^{2}}{\sigma^{2}_{v}}\bigg)\ell^{*}(\theta(t_{0})) \longrightarrow \chi^{2}_{1}, \quad \text{as}\,\, n \longrightarrow \infty.
\end{aligned}
\end{equation}
in distribution.
\end{theorem}
Thus, an asymptotic $(1-\alpha)100\%$ confidence interval for~$\theta(t)$~at a fixed time~$t$~is given as follows
\begin{equation}\label{}
\begin{aligned}
C(t)=\bigg\{\theta(t):\bigg(\dfrac{\sigma_{p}^{2}}{\sigma^{2}_{v}}\bigg)\ell^{*}(\theta(t))\le \chi^{2}_{1,\alpha}\bigg\},
\end{aligned}
\end{equation}
where $\chi^{2}_{1,\alpha}$ is the upper $\alpha-$quantile of the distribution of $\chi^{2}_1$.

\begin{proof} \hfill \break
Under the conditions of Theorem 2.1, \cite{yang2012} showed that 
\begin{equation*}
\begin{aligned}
\text{A1.}~& \dfrac{1}{n} \sum_{i=1}^{n} \widehat{W}_{i}^{2}(t) \xrightarrow[\hspace*{1cm}]{p} \sigma^{2}_{p}(t)\\
\text{A2.}~& \dfrac{1}{\sqrt{n}} \sum_{i=1}^{n} \widehat{W}_{i}(t) \xrightarrow[\hspace*{1cm}]{\mathcal{L}} N\big(0,\sigma^{2}_{v}(t)\big)
\end{aligned}
\end{equation*}

Following similar arguments as in the proof of Theorem 3.1 in \cite{rat2020} we have

\begin{equation}\label{eq22}
\begin{aligned}
\ell^{*}(\theta(t)) &= 2\sum_{i=1}^{n+1} \log \big(1 + \lambda(t) \widehat{W}_{i}(t)\big) \\
&=   2\sum_{i=1}^{n+1} \bigg\{\lambda(t) \widehat{W}_{i}(t) - \frac{\big(\lambda(t)  \widehat{W}_{i}(t)\big)^{2}}{2} \bigg \} + o_{p}(1).
\end{aligned}
\end{equation}
Using (\ref{eq22}), A1 and A2, we have

\begin{equation}\label{eq23}
\begin{aligned}
\bigg(\dfrac{\sigma_{p}^{2}}{\sigma^{2}_{v}}\bigg)\ell^{*}(\theta(t)) & = \bigg(\dfrac{\sigma_{p}^{2}}{\sigma^{2}_{v}}\bigg)  \sum_{i=1}^{n+1}\big(\lambda(t)  \widehat{W}_{i}(t)\big)^{2} + o_{p}(1) \\
 &= \dfrac{\bigg(\dfrac{1}{\sqrt{n}}\sum_{i=1}^{n+1}\lambda(t)  \widehat{W}_{i}(t)\bigg)^{2}}{\sigma_{v}^{2}}  ~~ \dfrac{\sigma^{2}_{p}}{\dfrac{1}{n}\sum_{i=1}^{n+1}\widehat{W}_{i}(t)^{2}}  + o_{p}(1) \\
& \xrightarrow[]{d}\chi^{2}_{1}.
\end{aligned}
\end{equation}
This completes the proof.
\end{proof}

\subsection{Transformed Empirical Likelihood for Generalized Lorenz Ordinate}
\cite{jing2017} proposed the transformed empirical likelihood (TEL) as a simple transformation of the original EL to tackle the under-coverage problem. They claimed that TEL is superior in small sample sizes and multidimensional situations. The transformed empirical log-likelihood ratio can be defined as
\begin{equation}\label{eq24}
\begin{aligned}
\ell_{t}\big(\theta(t), \gamma \big) = \ell(\theta(t))\times \max\bigg\{1 - \dfrac{\ell(\theta(t))}{n}, 1 - \gamma\bigg\},
\end{aligned}
\end{equation}
where $\ell(\theta(t))$ is given in (\ref{eqaug3022}) and $\gamma \in [0, 1]$. It should be noted that $\gamma = 1/2$ ensures the maximum expansion without violating the conditions (C2) stated in \cite{jing2017}. Hence, the transformed empirical log-likelihood ratio is defined as
\begin{equation}\label{eq25}
\begin{aligned}
\ell_{t}(\theta(t)) = \ell_{t}\bigg(\theta(t), \gamma = \dfrac{1}{2}\bigg) = \ell(\theta(t))\times \max\bigg\{1 - \dfrac{\ell(\theta(t))}{n}, ~~\dfrac{1}{2}\bigg\}.
\end{aligned}
\end{equation}
Thus, the transformed empirical log-likelihood ratio is given as

\begin{equation}
\ell_{t}(\theta(t)) =  \begin{cases}
      ~$$\ell(\theta(t))\bigg(1 - \dfrac{\ell(\theta(t))}{n} \bigg)$$\,  & \text{if   } $$\ell(\theta(t))\leq \dfrac{n}{2}$$, \\ \\
      ~$$$$\dfrac{\ell(\theta(t))}{2}$$  & \text{if   } $$\ell(\theta(t)) > \dfrac{n}{2}$$.$$
   \end{cases}
\end{equation}
Further, \cite{jing2017} showed that the TEL ratio meets four conditions that ensure the likelihood ratio's asymptotic properties. 

\begin{theorem}\label{thm32}
Assume that~$E(X^{2})<\infty,$~for all~$t_{0}\in (0,1),$ let $\ell_{t}(\theta(t_{0}))$ be the transformed log-empirical likelihood ratio function defined by (\ref{eq25}). We have
\begin{equation}\label{}
\begin{aligned}
\bigg(\dfrac{\sigma_{p}^{2}}{\sigma^{2}_{v}}\bigg)\ell_{t}(\theta(t_{0})) \longrightarrow \chi^{2}_{1}, \quad \text{as}\,\, n \longrightarrow \infty.
\end{aligned}
\end{equation}
in distribution.
\end{theorem}
Thus, an asymptotic $(1-\alpha)100\%$ confidence interval for~$\theta(t)$~at a fixed time~$t$~is given as follows
\begin{equation}\label{}
\begin{aligned}
C(t)=\bigg\{\theta(t):\bigg(\dfrac{\sigma_{p}^{2}}{\sigma^{2}_{v}}\bigg)\ell_{t}(\theta(t))\le \chi^{2}_{1,\alpha}\bigg\},
\end{aligned}
\end{equation}
where $\chi^{2}_{1,\alpha}$ is the upper $\alpha-$quantile of the distribution of $\chi^{2}_1$.

\begin{proof} \hfill \break
The proof is omitted as it is similar to the proof of Theorem 3.2 in \cite{rat2020}.
\end{proof}

\subsection{Transformed Adjusted Empirical Likelihood for Generalized Lorenz Ordinate}
\cite{pat2020} developed a hybrid method based upon AEL and TEL methods called transformed adjusted empirical likelihood (TAEL). The TAEL method combines the benefits of both AEL and TEL methods. Let $\gamma \in [0,1]$. We define

\begin{equation}\label{eq27}
\begin{aligned}
\ell_{t}^{*}\big(\ell^{*}(\theta(t)), \gamma\big) = \ell^{*}(\theta(t)) \times \max\bigg\{1 - \dfrac{\ell^{*}(\theta(t))}{n}, ~~1 - \gamma\bigg\},
\end{aligned}
\end{equation}
where $\ell^{*}(\cdot)$ defined in (\ref{eq12}). Thus, for $\gamma = 1/2$, the  transformed empirical log-likelihood ratio  $\ell_{t}^{*}(\theta(t))$ is defined as,
\begin{equation}\label{eq28}
\begin{aligned}
\ell_{t}^{*}\bigg(\ell^{*}(\theta(t)), \dfrac{1}{2}\bigg) = \ell^{*}(\theta(t)) \times \max\bigg\{1 - \dfrac{\ell^{*}(\theta(t))}{n},~~ \dfrac{1}{2}\bigg\}.
\end{aligned}
\end{equation}
This can be further viewed as
\begin{equation}\label{eqmay21}
       \ell_{t}^{*}(\theta(t)) =  \begin{cases}
      ~$$l^{*}(\theta(t))\bigg(1 - \dfrac{\ell^{*}(\theta(t))}{n} \bigg)$$\,  & \text{if  } $$\ell^{*}(\theta(t))\leq \dfrac{n}{2}$$, \\ \\
      ~$$$$\dfrac{\ell^{*}(\theta(t))}{2}$$  & \text{if   } $$\ell^{*}(\theta(t)) > \dfrac{n}{2}$$.$$
    \end{cases}
\end{equation}

\begin{theorem} \label{thm33}
  Assume that~$E(X^{2})<\infty,$~for all~$t_{0}\in (0,1),$ let $\ell_{t}^{*}(\theta(t_{0}))$ be the transformed log-empirical likelihood ratio function defined by (\ref{eqmay21}). We have
\begin{equation}\label{}
\begin{aligned}
\bigg(\dfrac{\sigma_{p}^{2}}{\sigma^{2}_{v}}\bigg)\ell_{t}^{*}(\theta(t_{0})) \longrightarrow \chi^{2}_{1}, \quad \text{as}\,\, n \longrightarrow \infty.
\end{aligned}
\end{equation}
in distribution.
\end{theorem}
Thus, an asymptotic $(1-\alpha)100\%$ confidence interval for~$\theta(t)$~at a fixed time~$t$~is given as follows
\begin{equation}\label{}
\begin{aligned}
C(t)=\bigg\{\theta(t):\bigg(\dfrac{\sigma_{p}^{2}}{\sigma^{2}_{v}}\bigg)\ell_{t}^{*}(\theta(t))\le \chi^{2}_{1,\alpha}\bigg\},
\end{aligned}
\end{equation}
where $\chi^{2}_{1,\alpha}$ is the upper $\alpha-$quantile of the distribution of $\chi^{2}_1$.
\begin{proof} \hfill \break
The proof of Theorem \ref{thm33} is similar to Theorem \ref{thm32}. In this case, the EL ratio defined in (\ref{eqaug3022}) is replaced by the adjusted empirical log-likelihood ratio defined in (\ref{eq12}). Thus, details are omitted here.
\end{proof}

\section{Simulation Study}
In this section, we conduct a simulation study to compare the performance of the proposed AEL, TEL, and TAEL-based confidence regions for the generalized Lorenz curve with EL-based confidence regions under various sample sizes in terms of coverage probabilities (CP) and mean lengths (ML) of the confidence intervals. The CP represents the proportion of times that the confidence regions contain the true value of the parameter among $N$ simulation runs. \\

Since most income distributions are positively skewed, the Weibull, Chi-square, and Skew-Normal distributions appear to provide a good fit for the income data. Thus, in our simulation study, we consider that the overall distribution function $F(x)$ is:
\begin{enumerate}
\item Weibull distribution with shape parameter $a = 1$, scale parameter $b = 2$. The pdf of the Weibull distribution is given by
\begin{equation*}\label{}
\begin{aligned}
f_{X}(x) = \bigg(\frac{a}{b} \bigg)\, \bigg(\frac{x}{b}\bigg)^{a-1} e^{-\big(x/b\big)^{a}},\quad x > 0
\end{aligned}
\end{equation*}
    \item $\chi^{2}$ distribution with $n= 3$ degrees of freedom. The pdf of the Chi-square distribution is given by
\begin{equation*}\label{}
\begin{aligned}
f_{X}(x) = \dfrac{1}{2^{n/2}\Gamma(n/2)}x^{n/2} e^{-x/2},\quad x >0
\end{aligned}
\end{equation*}
    \item Skew-Normal distribution with location parameter $\mu = 1$, scale parameter $\sigma = 3$, and shape parameter $\lambda = 5$. The pdf of the skew-normal distribution is given by:
\begin{equation*}\label{}
\begin{aligned}
f_{X}(x) = \frac{2}{\sigma}\phi\bigg(\frac{x-\mu}{\sigma}\bigg)\Phi\bigg(\lambda\frac{x-\mu}{\sigma}\bigg), \quad x \in \mathbb{R},
\end{aligned}
\end{equation*}
where $\phi(\cdot)$ and $\Phi(\cdot)$ are the pdf and cdf of the standard normal distribution. Further, in short-hand notation, we denote the skew-normal distribution by $X \sim SN(\mu,\sigma,\lambda)$.
\end{enumerate}

\begin{figure}[H]
  \centering
  \includegraphics[width=1\textwidth]{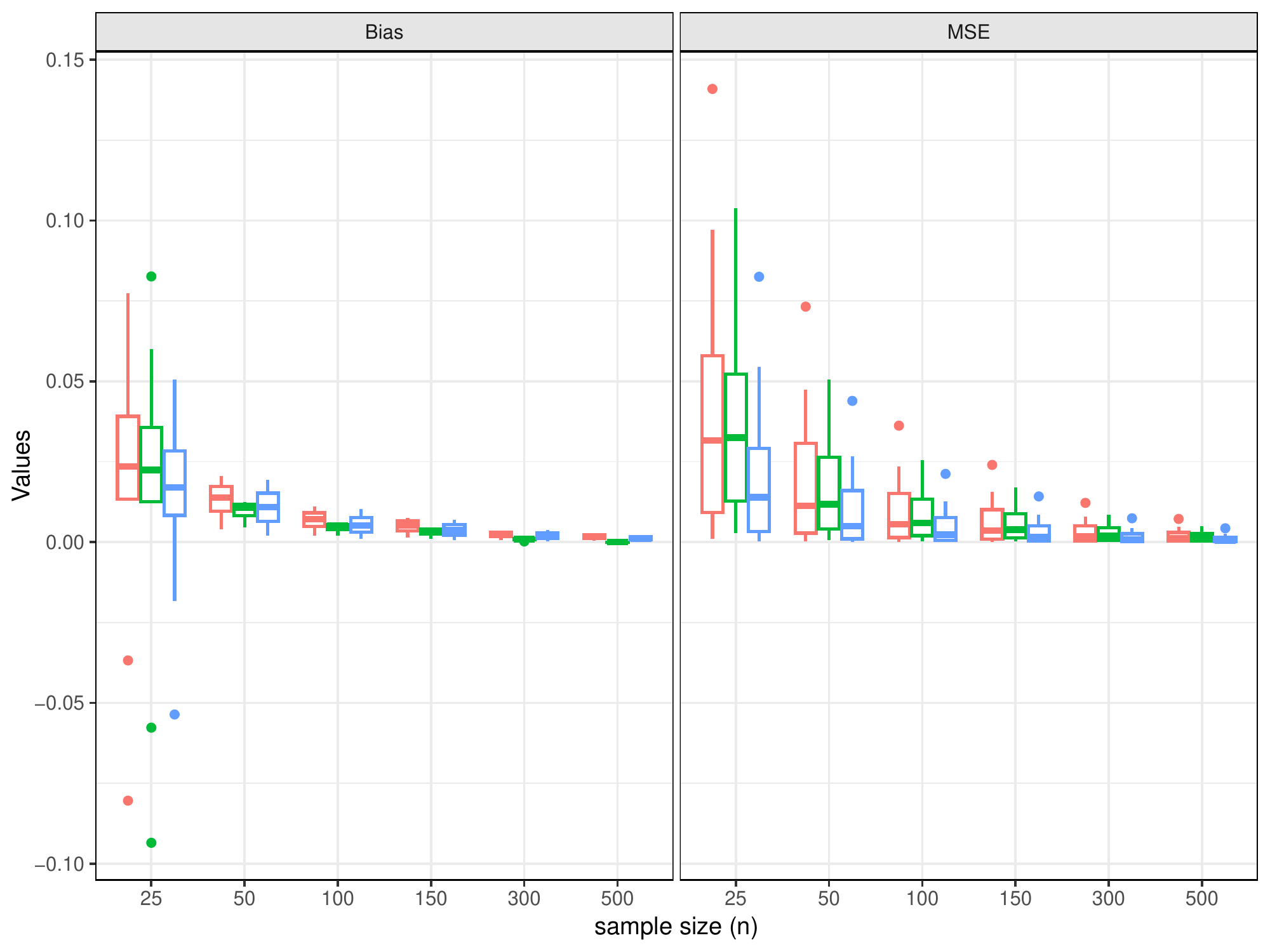}
  \caption{Bias and MSE of the generalized Lorenz ordinate of Weibull(1,2), $\chi^{2}_{3}$, and $SN(1,3,5)$ for various sample sizes}
  \label{fig0}
\end{figure}

We choose the sample size, $n = 25, 50, 100, 150, 300$ and $500$ representing a range from small to large, and values of $t_{0} = \{0.1, 0.2, 0.3, 0.4, 0.5, 0.6, 0.7, 0.8, 0.9$\}. Further, we set the nominal significance level $\alpha = 5\%$. The results are based on 10000 iterations. To assess the performance of the proposed methods, we consider two commonly used criteria for evaluating the goodness of a confidence interval procedure. These criteria are:
\begin{enumerate}
    \item Coverage probability: Preferably, close to 95\%.
    \item Mean lengths: Smaller is preferable.
\end{enumerate}
First, we compute bias and mean squared error (MSE) of the estimates for the generalized Lorenz ordinates for various distributions, including Weibull (1,2), $\chi_{3}^{2}$, and $SN(1,3,5)$. The results are summarized in Table \ref{table0} and graphed in Figure \ref{fig0}. It is evident that the bias of the estimate is consistently close to zero across all scenarios. Moreover, as the sample size increases, both bias and MSE generally decrease. Furthermore, it's notable that irrespective of the sample size, both bias and MSE increase as the value of $t$ increases.

\begin{figure}[H]
  \centering
  \includegraphics[width=1.05\textwidth]{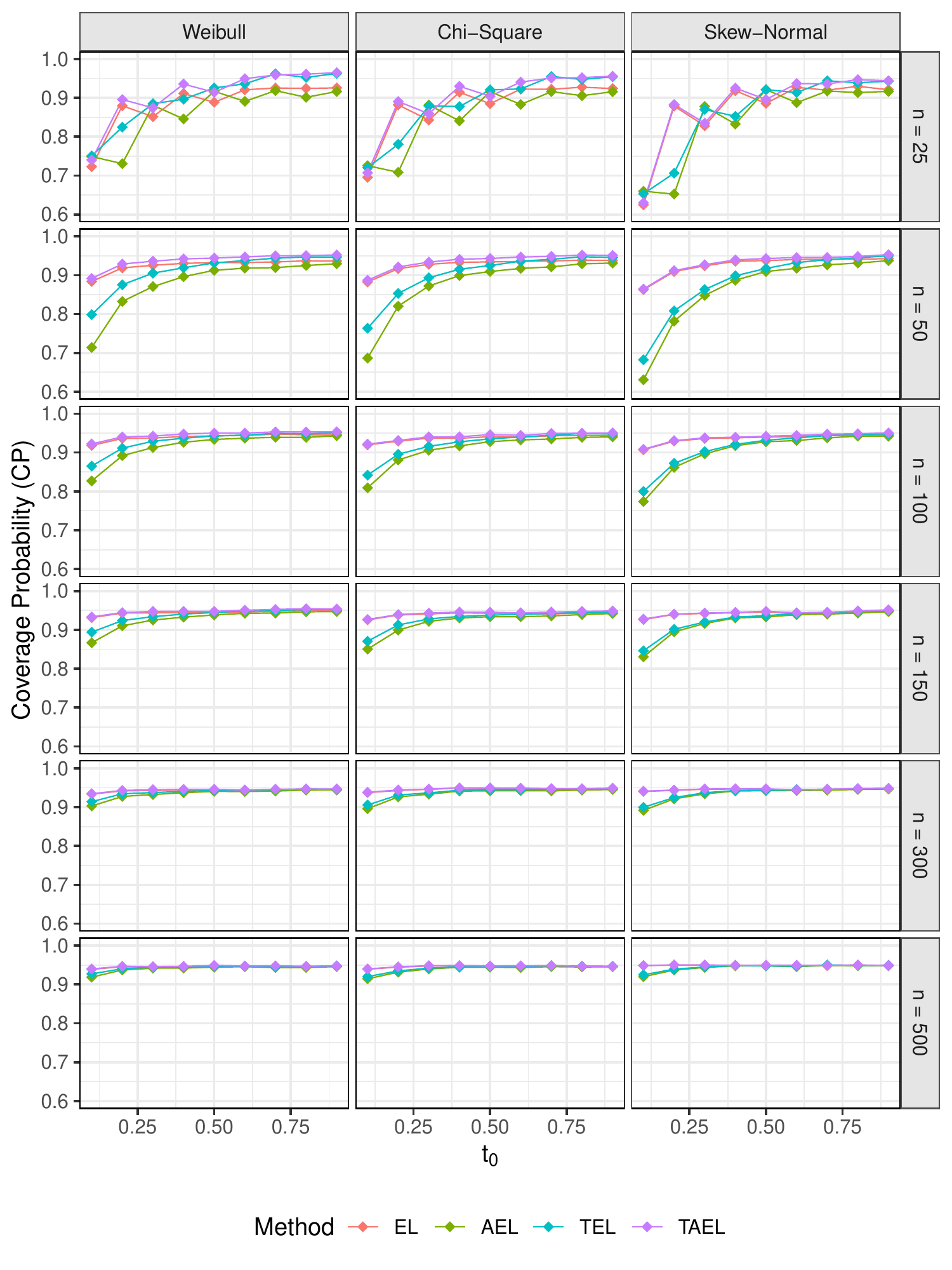}
  \caption{Coverage probabilities of the EL, AEL, TEL, and TAEL methods with a range of $t_{0}$ values  and various sample sizes for the generalized Lorenz ordinate of Weibull(1,2), $\chi^{2}_{3}$, and $SN(1,3,5)$}
  \label{fig1}
\end{figure}

\sloppy Next, we compute the coverage probability and mean lengths of the confidence regions for the generalized Lorenz ordinates $\theta(t)$.  The coverage probabilities are graphed in Figures \ref{fig1}. In all cases, the CP tends to increase as the sample size increases. Among all four methods, the TAEL method consistently provides the highest CP. In particular, the TAEL approach occasionally results in over-coverage issues. For $t \geq 0.5$, the TEL method outperforms EL, and AEL performs either slightly better or on par with the EL method. However, for $t < 0.5$, the EL approach performs significantly better than AEL and slightly better than the TEL method. \\

\begin{figure}[H]
  \centering
  \includegraphics[width=1.05\textwidth]{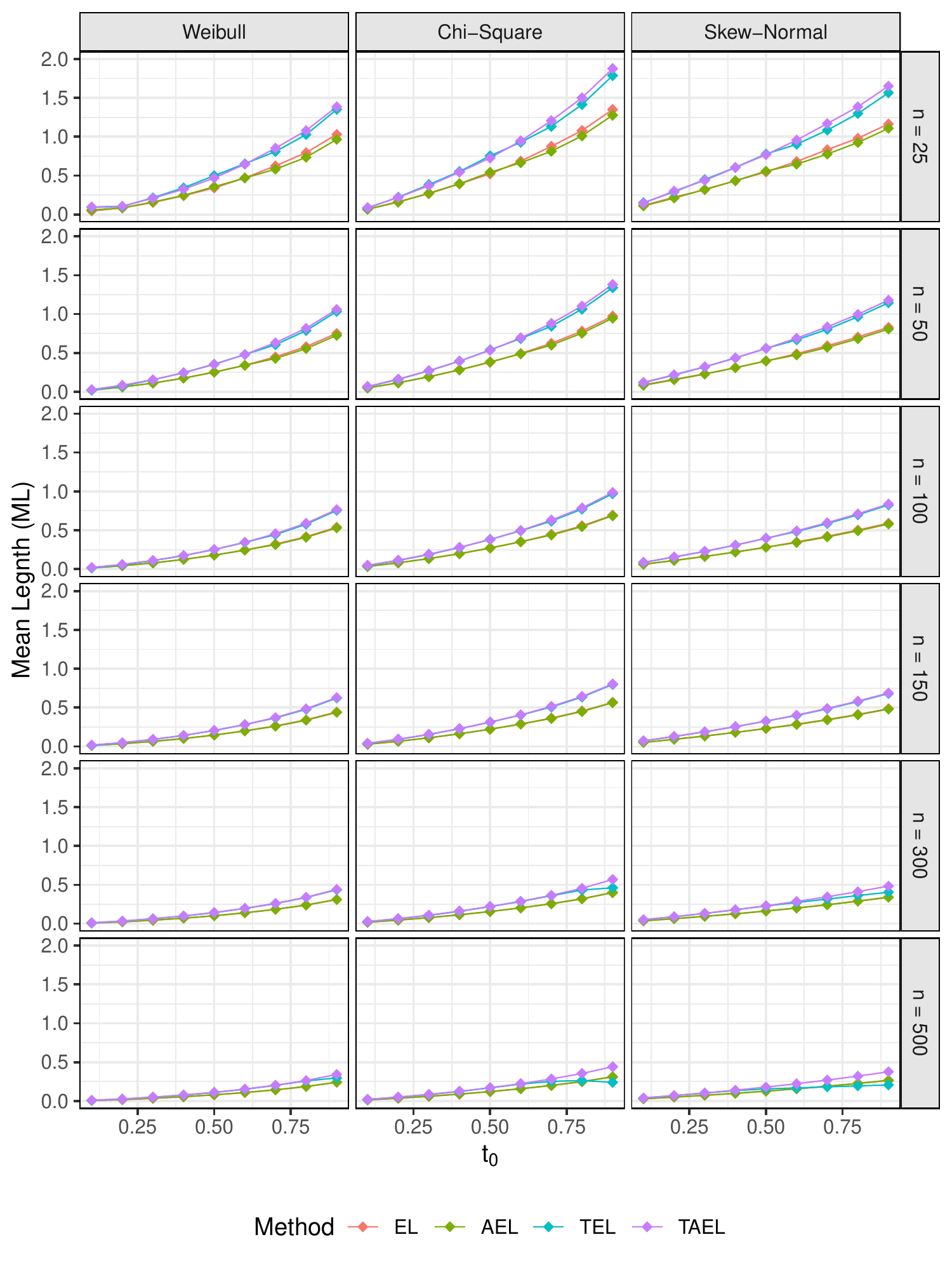}
  \caption{Mean lengths (ML) of the confidence intervals based on the EL, AEL, TEL, and TAEL methods with a range of $t_{0}$ values  and various sample sizes for the generalized Lorenz ordinate of Weibull(1,2), $\chi^{2}_{3}$, and $SN(1,3,5)$}
  \label{fig2}
\end{figure}

\sloppy When considering the mean lengths of confidence intervals, the TAEL method yields a slightly longer mean length, but it remains within an acceptable range. Among all four methods, the AEL results in the shortest confidence intervals. For $t<0.5$, the confidence intervals based on the TEL and TAEL approaches have approximately the same mean lengths. In addition, regardless of the method, as $t$ increases, the mean length also increases. However, as the sample size increases, the mean length of the confidence interval decreases. The mean lengths of confidence intervals are illustrated in Figure \ref{fig2}. \\

{\setlength{\tabcolsep}{0.8em}
\begin{table}[H]
\footnotesize
\caption{Bias and Mean Squared Error (MSE) for the Generalized Lorenz Curve with various probability distributions}
\centering
\begin{tabular}{cccccccccccccc}
\hline
  & & \multicolumn{2}{c}{Weibull(1,2)} & \multicolumn{2}{c}{$\chi_{3}^{2}$} & \multicolumn{2}{c}{$SN(1,3,5)$}    \\  \hline
\multirow{1}{*}{$n$}& \multirow{1}{*}{$t$} 	& Bias &  MSE & Bias &  MSE & Bias &  MSE  \\
\hline
25&	0.1&	0.0095&	0.0002&	0.0214&	0.0011&	0.0356&	0.0028\\
&	0.2&	0.0083&	0.0007&	0.0133&	0.0023&	0.0126&	0.0039\\
&	0.3&	0.0284&	0.0033&	0.0489&	0.0092&	0.0601&	0.0127\\
&	0.4&	0.0170&	0.0052&	0.0235&	0.0129&	0.0184&	0.0147\\
&	0.5&	0.0506&	0.0139&	0.0773&	0.0316&	0.0826&	0.0325\\
&	0.6&	0.0253&	0.0191&	0.0319&	0.0398&	0.0224&	0.0365\\
&	0.7&	-0.0184&	0.0291&	-0.0368&	0.0580&	-0.0577&	0.0522\\
&	0.8&	0.0328&	0.0545&	0.0391&	0.0972&	0.0236&	0.0744\\
&	0.9&	-0.0536&	0.0825&	-0.0804&	0.1409&	-0.0935&	0.1038\\\hline
							
50&	0.1&	0.0020&	0.0000&	0.0040&	0.0002&	0.0045&	0.0006\\
&	0.2&	0.0042&	0.0003&	0.0070&	0.0010&	0.0065&	0.0019\\
&	0.3&	0.0064&	0.0010&	0.0096&	0.0028&	0.0082&	0.0040\\
&	0.4&	0.0087&	0.0024&	0.0118&	0.0060&	0.0096&	0.0072\\
&	0.5&	0.0109&	0.0050&	0.0138&	0.0113&	0.0108&	0.0118\\
&	0.6&	0.0130&	0.0092&	0.0156&	0.0192&	0.0118&	0.0181\\
&	0.7&	0.0152&	0.0160&	0.0173&	0.0307&	0.0123&	0.0263\\
&	0.8&	0.0173&	0.0267&	0.0190&	0.0474&	0.0124&	0.0369\\
&	0.9&	0.0193&	0.0439&	0.0205&	0.0732&	0.0114&	0.0505\\\hline
							
100&	0.1&	0.0010&	0.0000&	0.0020&	0.0001&	0.0020&	0.0003\\
&	0.2&	0.0020&	0.0001&	0.0035&	0.0005&	0.0030&	0.0009\\
&	0.3&	0.0031&	0.0005&	0.0048&	0.0014&	0.0038&	0.0020\\
&	0.4&	0.0042&	0.0011&	0.0059&	0.0029&	0.0046&	0.0036\\
&	0.5&	0.0052&	0.0023&	0.0071&	0.0055&	0.0051&	0.0060\\
&	0.6&	0.0064&	0.0043&	0.0082&	0.0093&	0.0054&	0.0091\\
&	0.7&	0.0076&	0.0076&	0.0091&	0.0151&	0.0057&	0.0133\\
&	0.8&	0.0090&	0.0127&	0.0101&	0.0235&	0.0058&	0.0186\\
&	0.9&	0.0102&	0.0212&	0.0111&	0.0362&	0.0055&	0.0255\\\hline
							
150&	0.1	&0.0007&	0.0000&	0.0014&	0.0001&	0.0011&	0.0002\\
&	0.2&	0.0014&	0.0001&	0.0025&	0.0003&	0.0019&	0.0006\\
&	0.3&	0.0021&	0.0003&	0.0034&	0.0009&	0.0025&	0.0013\\
&	0.4&	0.0029&	0.0007&	0.0043&	0.0020&	0.0031&	0.0024\\
&	0.5&	0.0037&	0.0015&	0.0052&	0.0036&	0.0036&	0.0039\\
&	0.6&	0.0045&	0.0029&	0.0061&	0.0062&	0.0039&	0.0060\\
&	0.7&	0.0055&	0.0050&	0.0067&	0.0101&	0.0041&	0.0088\\
&	0.8&	0.0063&	0.0085&	0.0072&	0.0156&	0.0043&	0.0123\\
&	0.9	&0.0070&	0.0142&	0.0075&	0.0240&	0.0042&	0.0169\\\hline
							
300&	0.1&	0.0003&	0.0000&	0.0007&	0.0000&	0.0002&	0.0001\\
&	0.2	&0.0007&	0.0000&	0.0013&	0.0002&	0.0005&	0.0003\\
&	0.3&	0.0011&	0.0001&	0.0018&	0.0004&	0.0007&	0.0006\\
&	0.4&	0.0015&	0.0004&	0.0021&	0.0010&	0.0009&	0.0012\\
&	0.5&	0.0019&	0.0008&	0.0026&	0.0018&	0.0010&	0.0020\\
&	0.6&	0.0023&	0.0015&	0.0030&	0.0031&	0.0010&	0.0030\\
&	0.7&	0.0028&	0.0026&	0.0032&	0.0050&	0.0011&	0.0044\\
&	0.8&	0.0034&	0.0044&	0.0032&	0.0079&	0.0010&	0.0062\\
&	0.9&	0.0038&	0.0074&	0.0030&	0.0122&	0.0008&	0.0085\\\hline
							
500&	0.1&	0.0002&	0.0000&	0.0005&	0.0000&	-0.0002&	0.0001\\
&	0.2&	0.0004&	0.0000&	0.0009&	0.0001&	-0.0001&	0.0002\\
&	0.3&	0.0007&	0.0001&	0.0012&	0.0003&	0.0000&	0.0004\\
&	0.4&	0.0009&	0.0002&	0.0015&	0.0006&	0.0000&	0.0007\\
&	0.5&	0.0011&	0.0005&	0.0018&	0.0011&	0.0000&	0.0012\\
&	0.6	&0.0014&	0.0009&	0.0021&	0.0019&	0.0000&	0.0018\\
&	0.7&	0.0016&	0.0015&	0.0023&	0.0030&	0.0000&	0.0026\\
&	0.8&	0.0019&	0.0026&	0.0024&	0.0047&	-0.0001&	0.0037\\
&	0.9&	0.0022&	0.0043&	0.0024&	0.0072&	-0.0002	&0.0050\\ \hline

\end{tabular}
\label{table0}
\end{table}}

\section{Application to Real Data}

\begin{figure}[H]
  \centering
  \includegraphics[width=1\textwidth]{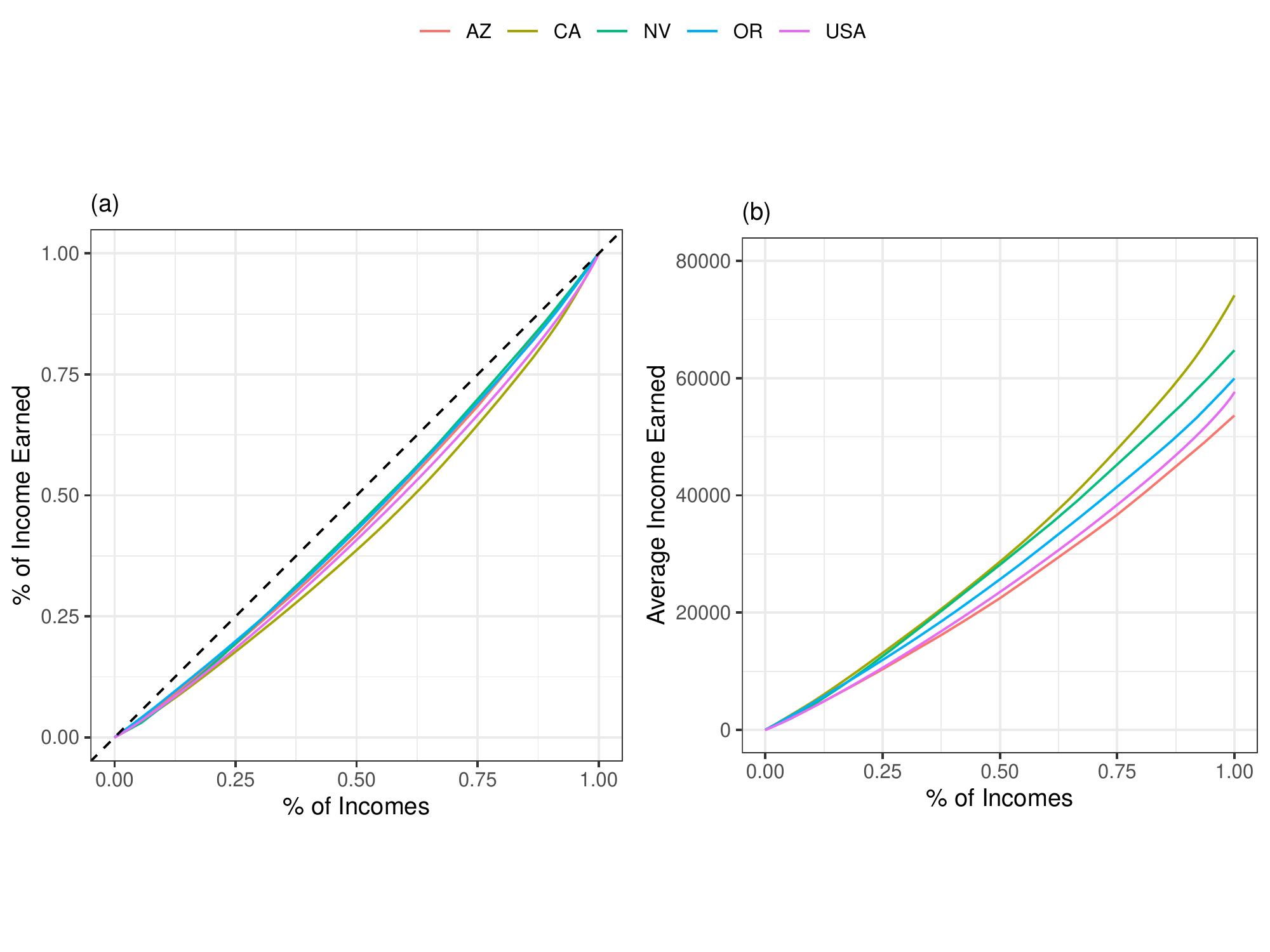}
  \caption{(a) Lorenz Curves and (b) Generalized Lorenz Curves for Arizona (AZ), California (CA), Nevada (NV), Oregon (OR), and USA}
  \label{fig3}
\end{figure}

\begin{figure}[H]
  \centering
  \includegraphics[width=1\textwidth]{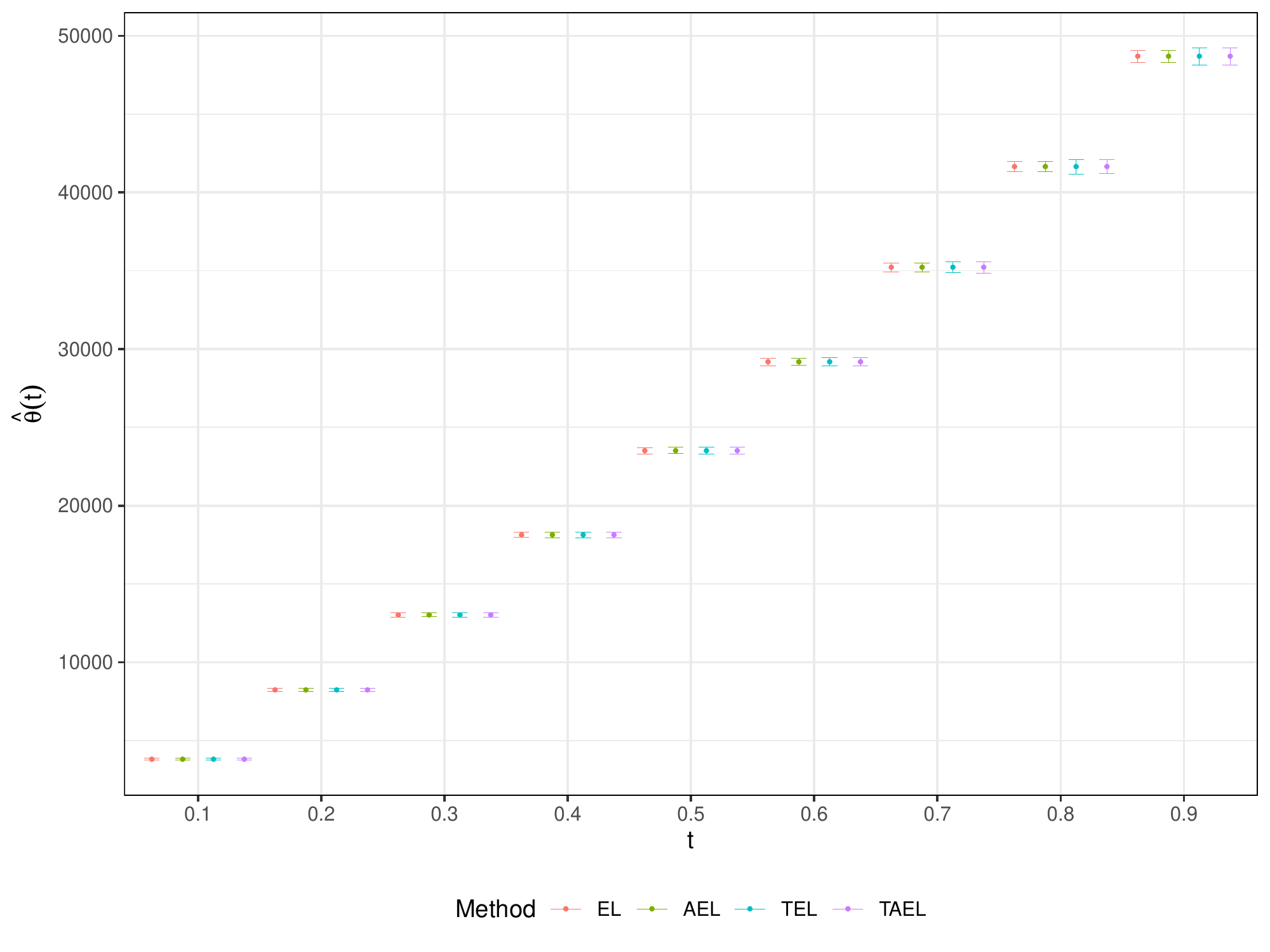}
  \caption{A 95\% confidence interval for the generalized Lorenz ordinate based on EL, AEL, TEL, and TAEL methods with various $t$ values}
  \label{fig31}
\end{figure}

In this section, we demonstrate the effectiveness of the proposed AEL, TEL, and TAEL methods for generalized Lorenz ordinate by constructing confidence intervals for Median Household Income in 2020. The data set is available \url{https://www.ers.usda.gov/data-products/county-level-data-sets/download-data/}, which contains 3194 observations of the median household income in 2020, and they are grouped by state or county name. We mainly focus on examining the median income distribution of households in Arizona (AZ), California (CA), Nevada (NV), Oregon (OR), and the US as a whole. The Lorenz curves and the generalized Lorenz curves for the four states and the US are graphed in Figure \ref{fig3}. The black dashed line represents the equality line. It is evident that Arizona has the Lorenz curve that is closest to the equality line. Further, the USA has the most unequal income distribution, followed by California and Nevada.  We also compute a 95\% confidence interval for the generalized Lorenz ordinate using EL, AEL, TEL, and TAEL methods with various $t$ values considering $t = \{0.1, 0.2, 0.3, 0.4, 0.5, 0.6, 0.7, 0.8, 0.9\}$. We considered all 3194 observations for the purposes of this analysis. The results are summarized in Table \ref{table4} and they are plotted in Figure \ref{fig31}. We notice that EL and AEL perform similarly while TEL and TAEL roughly produce the same confidence intervals. Additionally, the AEL approach consistently yields a shorter confidence length than the other three methods.

{\setlength{\tabcolsep}{1.1em}
\begin{table}[H]
\footnotesize
\caption{A 95\% confidence intervals for the generalized Lorenz ordinates with various values of $t$ for Median Household Income in 2020 in America. }
\centering
\begin{tabular}{cccccccccccccc}
\hline
\multirow{1}{*}{$t$}& $\hat{\theta}(t)$ & \multirow{1}{*}{Method} 	& Lower &  Upper & Length  \\
\hline
0.1& 3815.195&	EL&	3759.1760&	3871.7118&	112.5535\\
&&	AEL&	3757.9974&	3870.5021&	112.5048\\
&&	TEL&	3756.8501&	3871.6701&	114.8200\\
&&	TAEL&	3758.0284&	3872.8800&	114.8516\\ \hline
0.2& 8240.058&	EL&	8145.7837&	8334.8855&	189.1324\\
&&	AEL&	8143.2260&	8332.2851&	189.0591\\
&&	TEL&	8140.5307&	8335.0126&	194.4819\\
&&	TAEL&	8143.0881&	8337.6133&	194.5251\\\hline
0.3& 13027.020 &	EL&	12896.9940&	13157.5745&	260.6236\\
&&	AEL&	12908.4210&	13168.9741&	260.5531\\
&&	TEL&	12903.4394&	13173.9962&	270.5568\\
&&	TAEL&	12892.0084&	13162.6007&	270.5923\\\hline
0.4	& 18138.190 & EL&	17973.2271&	18303.5235&	330.3522\\
&&	AEL&	17967.5732&	18297.8236&	330.2504\\
&&	TEL	&   17959.0887&	18306.3482&	347.2595\\
&&	TAEL&	17964.7429&	18312.0477&	347.3048\\\hline
0.5& 23514.140&	EL&	23312.6016&	23715.7678&	403.2358\\
&&	AEL&	23322.5887&	23725.6854&	403.0967\\
&&	TEL&	23307.9250&	23740.3629&	432.4379\\
&&	TAEL&	23297.9302&	23730.4531&	432.5230\\\hline
0.6& 29186.760 &	EL&	28947.5843&	29425.4732&	477.9736\\
&&	AEL&	28956.5766&	29434.2810&	477.7044\\
&&	TEL	&   28930.0349&	29460.7166&	530.6816\\
&&	TAEL&	28921.0092&	29451.9423&	530.9331\\\hline
0.7	& 35224.900 & EL&	34941.7069&	35506.5735&	564.9692\\
&&	AEL&	34929.3294&	35494.0013&	564.6719\\
&&	TEL&	34863.9490&	35558.6065&	694.6575\\
&&	TAEL&	34876.1887&	35571.3135&	695.1248\\\hline
0.8& 41650.550 &	EL&	41323.3627&	41974.1675&	650.9197\\
&&	AEL&	41328.4634&	41979.0655&	650.6021\\
&&	TEL	&   41191.9480	&42112.0198&	920.0719\\
&&	TAEL&	41186.8051&	42107.1635&	920.3584\\\hline
0.9& 48694.690 &	EL&	48301.5894&	49079.4584&	777.9221\\
&&	AEL&	48304.5085&	49082.1298&	777.6213\\
&&	TEL	&   48139.2496&	49239.0692&	1099.8196\\
&&	TAEL&	48136.2785&	49236.4482&	1100.1697\\ \hline

\end{tabular}
\label{table4}
\end{table}}

\section{Discussions}

In this article, we proposed powerful nonparametric EL-based methods for constructing confidence intervals for generalized Lorenz ordinate. These methods include the adjusted empirical likelihood (AEL), the transformed empirical likelihood (TEL), and the transformed adjusted empirical likelihood (TAEL). We derive the limiting distributions of the generalized Lorenz ordinate based on the AEL, TEL, and TAEL methods.  Simulations show that the proposed TEL and TAEL methods improve the coverage probability compared to the EL method. According to the simulation study, we highly recommend the TAEL method for $t<0.5$ and small sample sizes. When $t \geq 0.5$, both the EL and AEL approaches yield comparable results for medium and large samples, making AEL an additional option. While the TEL method is suitable for large samples $(n\geq 300)$, the TAEL method is appropriate for all sample sizes. In real-world applications, we recommend the TAEL approach as it consistently offers superior coverage compared to the other three methods. It's worth noting although the confidence intervals based on the TAEL approach are longer than others, they remain within an acceptable range. Our real-world data application demonstrates that the proposed methods are competitive with the EL method while also addressing its limitations.

\section*{Declarations}

\noindent
\textbf{Conflict of interest }The authors declare that they have no conflict of interest. \\

\noindent

\section*{Acknowledgements}
We would like to thank two anonymous referees for their comments, which have contributed to this improved version of the work. We also would like to express our appreciation to the Office of Student Research (OSR) at California State University, San Bernardino, for creating a supportive environment for conducting this research.

\section*{References}
\renewcommand{\section}[2]{}
\bibliographystyle{apalike}
\bibliography{References}
\vspace{2cm}

\end{document}